\documentclass[conference]{IEEEtran}
\IEEEoverridecommandlockouts
\usepackage{cite}
\usepackage{amsmath,amssymb,amsfonts}
\usepackage{algorithmic}
\usepackage{graphicx}
\usepackage{textcomp}
\usepackage{xcolor}
\usepackage{subfigure}

\def\BibTeX{{\rm B\kern-.05em{\sc i\kern-.025em b}\kern-.08em
    T\kern-.1667em\lower.7ex\hbox{E}\kern-.125emX}}
\begin{document}

\title{Network Coding for Critical Infrastructure Networks}

\author{\IEEEauthorblockN{Rakesh Kumar\IEEEauthorrefmark{1},
Vignesh Babu\IEEEauthorrefmark{2} and
David M. Nicol \IEEEauthorrefmark{3}}
\IEEEauthorblockA{Information Trust Institute,\\
University of Illinois, Urbana-Champaign\\
Urbana, Illinois\\
Email: \IEEEauthorrefmark{1}kumar19@illinois.edu,
\IEEEauthorrefmark{2}babu3@illinois.edu,
\IEEEauthorrefmark{3}dmnicol@illinois.edu}}

\maketitle

\begin{abstract}
The applications in the critical infrastructure systems pose simultaneous resilience and performance requirements to the underlying computer network. To meet such requirements, the networks that use the store-and-forward paradigm poses stringent conditions on the redundancy in the network topology and results in problems that becoming computationally challenging to solve at scale. However, with the advent of programmable data-planes, it is now possible to use linear network coding (NC) at the intermediate network nodes (i.e. hardware and software switches) to meet resilience requirements of the applications. To that end, we propose an architecture that realizes linear NC in programmable networks by decomposing the linear NC functions into the atomic coding primitives. We designed and implemented the primitives using the features offered by the P4 ecosystem. Using an empirical evaluation of an open-source prototype, we show that the theoretical gains promised by linear network coding can be realized with a per-packet processing cost. 
\end{abstract}

\begin{IEEEkeywords}
Resilience, Network Coding, Software Defined Networking
\end{IEEEkeywords}

\section{Introduction}
\label{sec:introduction}

The applications that constitute the critical infrastructure (e.g. smart power generation and distribution systems, oil refineries etc.) have a unique set of requirements regarding their underlying communication networks. For example, such applications require that their communication is \textit{seamlessly} resilient against link or device failures. Furthermore, these applications also require a \textit{predictable} end-to-end delay for data delivery in multicast settings \cite{bakken2011smart} \cite{leggett2010station}. Such resilience and performance requirements cannot be simultaneously accomplished by mere over-provisioning of network resources such as topological redundancy or bandwidth. 

Rather, in the packet store-and-forward paradigm, the resiliency is provided by carefully routing the packets around a failed link or network device \cite{elhourani2016ip} \cite{pignolet2017load} \cite{chiesa2016quest} \cite{chiesa2016resiliency}. However, such an approach requires solving complex combinatorial problems. Similarly, performance requirements (e.g. end-to-end delay) are met by solving resource allocation problems on a per-flow basis \cite{kumar2017end}, but even a static resource allocation for flows that have such performance requirements is an NP-complete problem \cite{wang1996quality}. Therefore, combining the performance and resilience requirements poses an intractable problem. 

Such intractability is a result of \textit{hard} routing and resource allocation decisions that are in turn a consequence of the atomic nature of a packet flow in the store-and-forward paradigm. In this paradigm, a flow has to originate at a source port and follow a specific path to arrive at the destination(s) without any modifications to its contents. However, network coding converts this hard decision into one of many \textit{soft} decisions by mixing packets at intermediate network devices using algebraic coding. In theory, NC promises to provide seamless resilience to failures for critical infrastructure applications over the store-and-forward paradigm \cite{ahlswede2000network} \cite{koetter2003algebraic}. However, practical NC that achieves the promised theoretical gains has remained elusive.

Clearly, NC is realized when the intermediate network devices can be programmed to implement the packet coding and decoding capabilities. While there have been successful attempts to demonstrate the efficacy of using inter-session NC in wireless networks \cite{katti2008xors} \cite{ostovari2014network}, the progress on the widespread adoption of the same has been disappointing. In part, the reason has been the practical issues of retrofitting NC onto the prevalent networking architecture. These issues have been addressed in various ingenious efforts in the past \cite{sundararajan2009network} \cite{chou2003practical}. But, more importantly, the adoption of NC has been stifled due to a lack of programmable platforms that can implement novel data-plane methods at scale. Historically, the switch ASIC architectures that implement data-plane functionality have been optimized for ever-increasing line-speed performance at the expense of programmability. However, very recently, with the advent of programmable data-planes \cite{bosshart2014p4}, it has become possible to not only experiment \cite{wang2017p4fpga}, but also deploy new network functions using a flexible data-plane architecture in production networks \cite{tofino}. 

Based on these developments, we devise an architecture capable of simultaneously meeting resilience and performance requirements of the data streams generated by applications in critical infrastructure systems. To that end, we present one that leverages programmable networks to replace routing algorithms with NC functions. Our contributions include:

\begin{itemize}
\item A library of atomic network coding primitives implemented using the programmable data-planes. 

\item Use of the proposed primitives to construct linear network coding functions capable of achieving specific requirements for applications' data streams.

\item Evaluation of the coding functions to show that the seamless resilience and multicast rate gains are obtained at a small per-packet processing cost of coding and decoding the packets in the data-plane.
\end{itemize}

The remainder of this paper is organized as follows: Section \ref{sec:related_work} discusses related work; Section \ref{sec:background} discusses background and how the features of programmable data-planes affect the design of coding functions and primitives; Section \ref{sec:architecture} proposes an architecture to implement coding functions; Section \ref{sec:design} discusses the design of various elements of the proposed architecture; Section \ref{sec:evaluation} evaluates the performance and costs of using the proposed design; and Section \ref{sec:conclusion_future_work} concludes and discusses future work.
\section{Related Work}
\label{sec:related_work}

There has been prior work in the store and forward paradigm that allows nearly instantaneous failure recovery. When such failures are addressed reactively, they result in prohibitively large restoration time for critical infrastructure applications \cite{van2014fast} \cite{bakken2011smart}. There are proactive approaches to deal with such failures which use the mechanisms local to a switch to reroute traffic on an alternative path \cite{elhourani2016ip} \cite{pignolet2017load} \cite{chiesa2016quest} \cite{chiesa2016resiliency}. However, such approaches require $k$-connected network topologies for sustaining $k$ link failures, thus incurring a large overhead in procuring and maintaining such networks. Furthermore, these approaches require solving combinatorial problems to choose alternative links in the event of link failures. These approaches also lead to new problems such as the need to "load-balance" resilience so that a small set of links does not become too critical for the resulting network after the failures.

Recently, in order to meet the performance guarantees in store-and-forward networks, the standard bodies have proposed standards for special-purpose hardware \cite{tsn} \cite{teener2013heterogeneous}. However, using such special hardware incurs large capital and recurring expenses. To that end, some recent work has proposed mechanisms to simultaneously meet per-flow end-to-end delay and bandwidth requirements using software-defined commodity networks \cite{kumar2017end}. However, this work solves a resource allocation problem using a heuristic for a multi-constraint path problem that provides no guarantees of optimality.

The seminal work that demonstrated a practical mechanism to implement NC by using simulations was done by Chou et. al. \cite{chou2003practical}. This work focused on coding batches of data which is incompatible with the acknowledgement mechanisms of TCP. Subsequently, there has been work that demonstrated TCP throughput gains with the use of NC \cite{sundararajan2009network} \cite{krigslund2015network} with deployable implementations. However, these efforts focus on intra-session coding at source only and the goodput gains obtained due to intermittent packet loss.

There have been several prior efforts to implement NC top of on the application layer, either in an overlay topology \cite{gkantsidis2006comprehensive} \cite{gkantsidis2005network} \cite{wu2008dynamic} or as a virtual network function \cite{zhang2017virtualized}. While implementing NC in the application layer offers flexibility and variety in the type applications that can be materialized, the cost of taking packets from the network interface and processing them in upper layers can be high and can be mitigated by implementing NC in the data-plane of network devices. 

Finally, COPE \cite{katti2008xors} demonstrated the benefits of inter-session coding in the specific setting of wireless networks by utilizing the broadcast property of the media with a clever heuristic. The types of benefits that COPE extracted using a specially designed architecture can now be replicated for wired networks by implementing NC using standard platforms such as the P4 ecosystem.
\section{Motivation and Background}
\label{sec:background}

\subsection{Why use Network Coding?}
In the store-and-forward paradigm, each flow originates at a source port and follows a deterministic path to arrive at the destination port(s), however, its contents are immutable during the transit. Hence, due to immutability, when globally optimal decisions for resource allocation for flows are to be made, the resources are allocated separately for each flow at the network devices. Furthermore, in the event of a link or device failure, the flows must be routed around the failure entirely. Due to these requirements of immutability, solving for performance guarantees and resilience requirements results in formulation of problems that are intractable \cite{wang1996quality} \cite{kumar2017end} or combinatorially complex \cite{elhourani2016ip} \cite{pignolet2017load} \cite{chiesa2016quest} \cite{chiesa2016resiliency}. 

The NC paradigm approaches the problem of delivering data from point A to point B by allowing intermediate nodes within the network to code and recode the packets. This paradigm has many promising theoretical properties. For example, in their seminal work, Ahlswede et. al. \cite{ahlswede2000network} showed that, given a network represented as a multigraph $G(V,E)$, network coding can enable a sender $s \in V$ to communicate with a set of receivers $T \subset V\setminus s$, at a multicast rate equal to the minimum max-flow from the sender to any of the $t \in T$. Li et. al. \cite{li2003linear} showed that using linear codes on the network nodes are sufficient to achieve this rate. Koetter and Medard \cite{koetter2003algebraic} extended the theorem to the cases when the edges in $E$ are subject to failures and showed that the linear codes can achieve minimum max-flow even after failures. Finally, Ho et. al. \cite{ho2006random} showed that the random linear network codes suffice to achieve the same. 

\subsection{Why use the P4 ecosystem?}
The implementation of linear NC requires two types of computations: First, there are network-level operations such as computation of coding co-efficients or designation of various roles to the individual nodes based on the topology and the application requirements. These operations can be performed in the programmable control-plane. Second, there is the simple arithmetic operations (e.g. addition, multiplication) that are performed on an individual or a small batch of packets in the applications' data stream. In order for coding to scale to line-speeds, these operations have to be performed on the individual network devices using a programmable data-plane architecture. 

The P4 ecosystem is a programmable data-plane architecture has been gaining traction in both academia and industry for implementing novel data-plane functions. It comprises an open-source $P4_{16}$ language \cite{bosshart2014p4} \cite{p4_16_spec} and the accompanying Portable Switch Architecture (PSA) \cite{p4_16_psa}. The ecosystem has accelerated the design and adoption of novel network functions by enabling fully programmable data-planes without compromising the line-speed performance of modern network devices. 

The primary goal of the P4 ecosystem is to make data-planes programmable by allowing expression of per-packet computations performed on a network device. As such, it is not designed to enable network coding applications. However, it does provide several features that make it well-suited as a platform for implementing linear NC functions for failure resilience and multicast rate enhancements. Below, we briefly describe some of the relevant features:

\begin{itemize}
\item \textbf{Customizable Packet Processing Pipelines}: There are separate ingress and egress pipelines on each PSA device which can be configured from the control-plane. These pipelines are constructed using tables. Each table has a set of fields (called its \texttt{key}) that determines the packets that \textit{match} it. Each table also has associated C-link sub-routines called actions. The actions can perform nearly arbitrary operations on the packet headers including for example addition, multiplication and XOR.

\item \textbf{Packet Cloning \& Recirculation}: Cloning makes copies of packets on the egress pipeline, while recirculation sends the packets from the egress pipeline to the ingress pipeline. Both of these features can be used in tandem to \textit{generate} a new packet for coding/decoding operations. Furthermore, since P4 does not have a primitive analogous to a loop in imperative programming languages, packet cloning and recirculation can also be used to create one without any intervention from end-hosts.

\item \textbf{Registers}: Registers are essentially global variables that can hold global state independent of any specific packet in any given pipeline. These registers can be used to drive state machines and implement data structures that hold packets that are required to be coded.

\item \textbf{Extensibility}: P4 allows extension of the core language by using a construct called \texttt{extern}. Essentially, this construct allows another level of flexibility to implement features that do not exist in the language. Such flexibility may be crucial to any implementation of NC that goes beyond simple linear codes.
\end{itemize}
\section{Architecture}
\label{sec:architecture}

\begin{figure}
\centering
\includegraphics[width=3.5in]{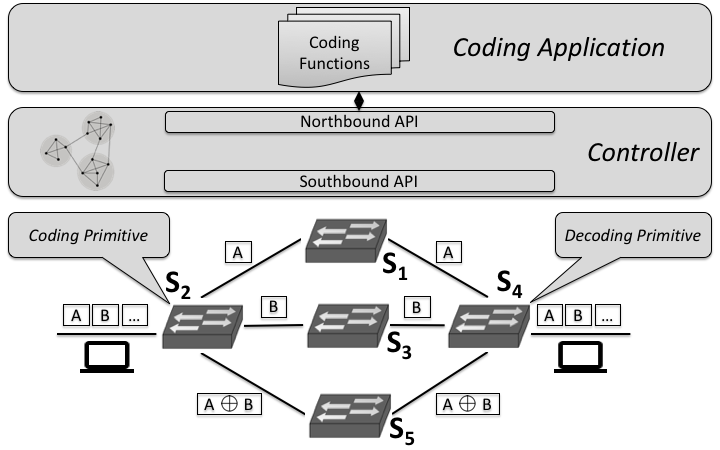}
\caption{Architecture}
\label{fig:architecture}
\end{figure} 

The Figure \ref{fig:architecture} illustrates the proposed architecture which enables implementation of linear network coding using P4 devices. This particular example shows a steam of packets carrying applications' data originates at the host on the left side and terminates on the host at the right side. Fundamentally, we assume that a packet stream can be divided into a batches of packets. These batches are then processed by individual devices to achieve the NC gains. 

We define a \textit{coding function} as the realization of a linear code to improve resilience or throughput of a unicast/multicast data stream. For example, in Figure \ref{fig:architecture}, the function implements a diversity code \cite{ayanoglu1993diversity} to provide resilience to failure of any one of the three paths between $S_1$ and $S_3$. The function replaces IP forwarding and spans one or more P4 enabled devices. A northbound coding application implements multiple coding functions that operate simultaneously across the network. 

A \textit{coding primitive} is an atomic block of functionality implemented on the individual P4 switches. For example, in Figure \ref{fig:architecture}, switches $S_1$ and $S_3$ implement the coding and decoding primitives respectively. Each primitive operates independently of the others. Each incident stream of packets on the device is subject to one or more primitives. A switch can process multiple data streams simultaneously. Each switch's configuration contains the identifier for the data streams and the exact sequence of coding primitives applied to each of them.

\section{Design}
In this section, we first describe our design for the coding functions in the control-plane. Next, we discuss the packet header that is used to coordinate primitives for a given coding function and finally describe how our design of the coding primitives in the data-plane using the P4 ecosystem. 

\label{sec:design}

\begin{figure*}[ht!]
\centerline{\subfigure{\includegraphics[width=3.5in]{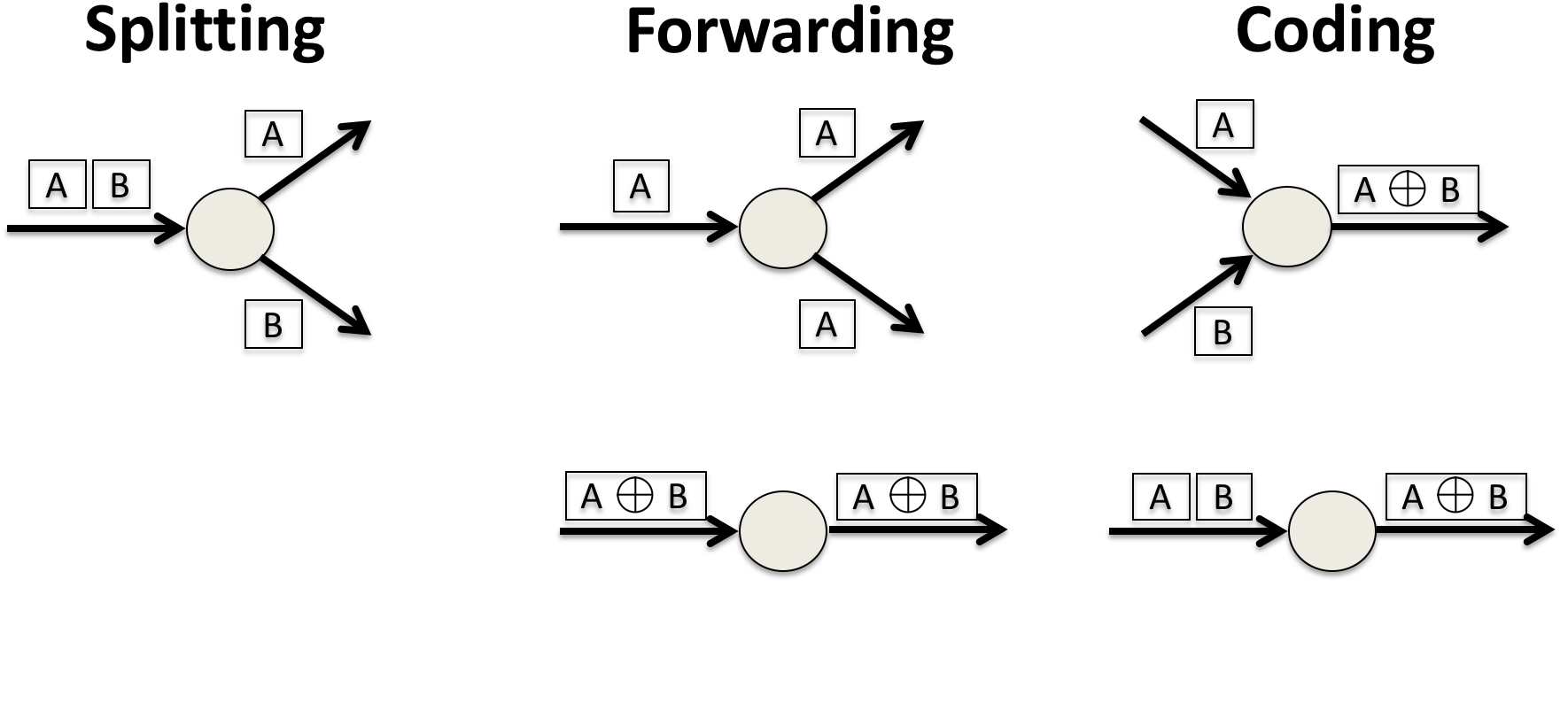}}
\subfigure{\includegraphics[width=2.6in]{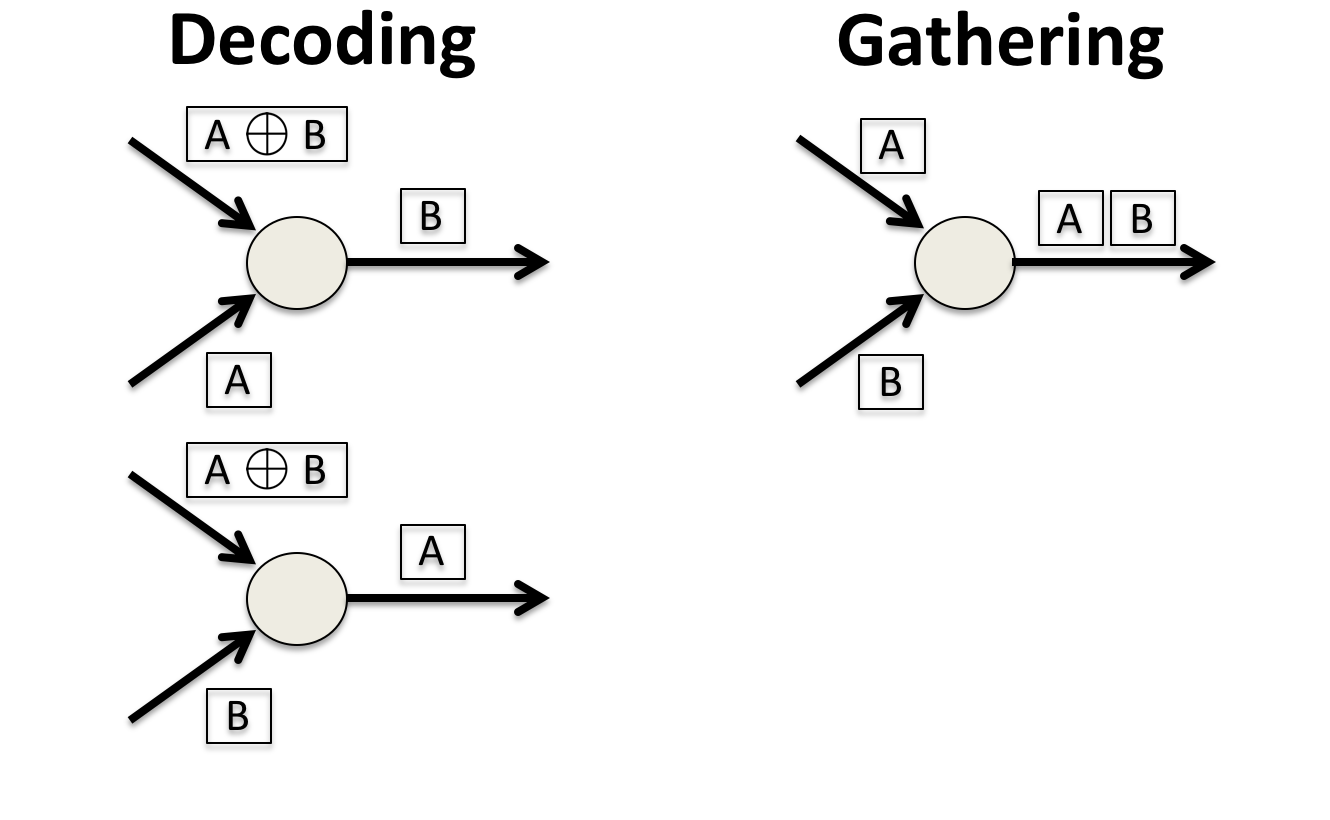}}\hfil} 
\caption{Network Coding Primitives}
\label{fig:primitives}
\end{figure*}

\begin{figure*}[ht!]
\centerline{\subfigure(a){\includegraphics[width=3.8in]{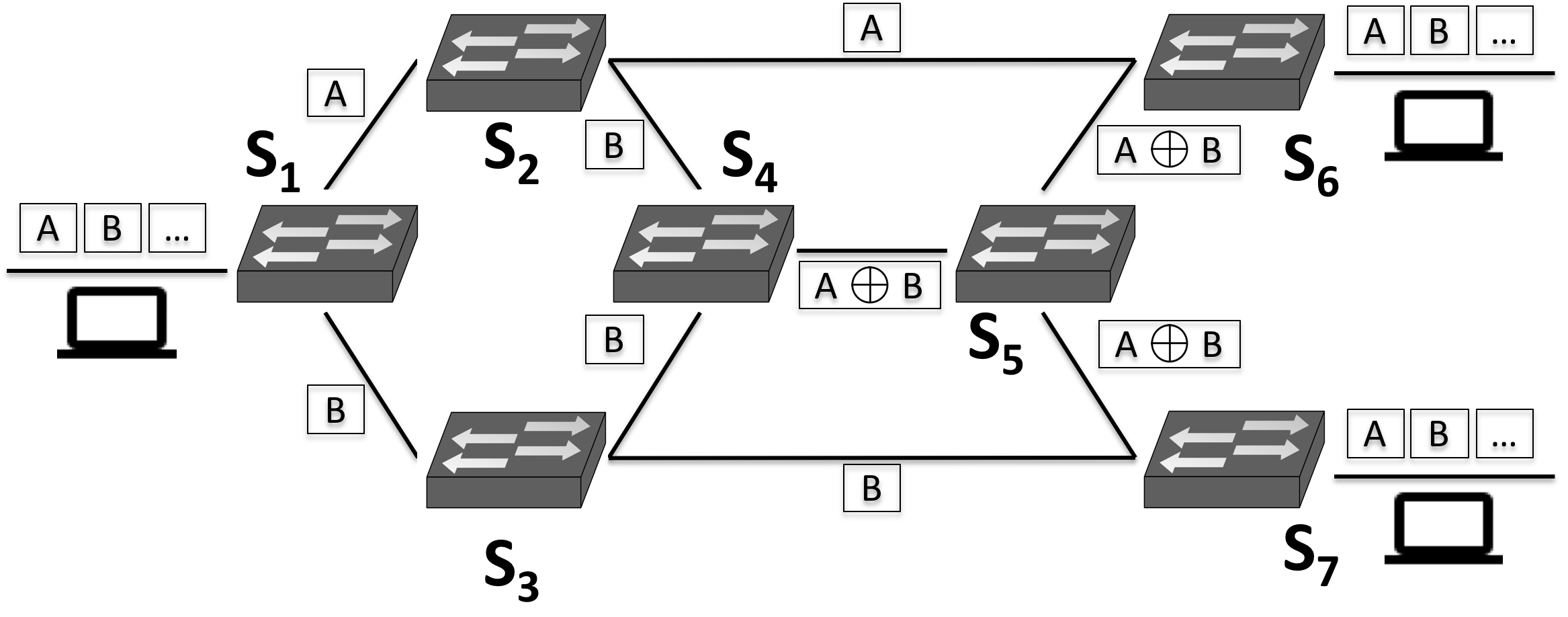}}
\subfigure(b){\includegraphics[width=3in]{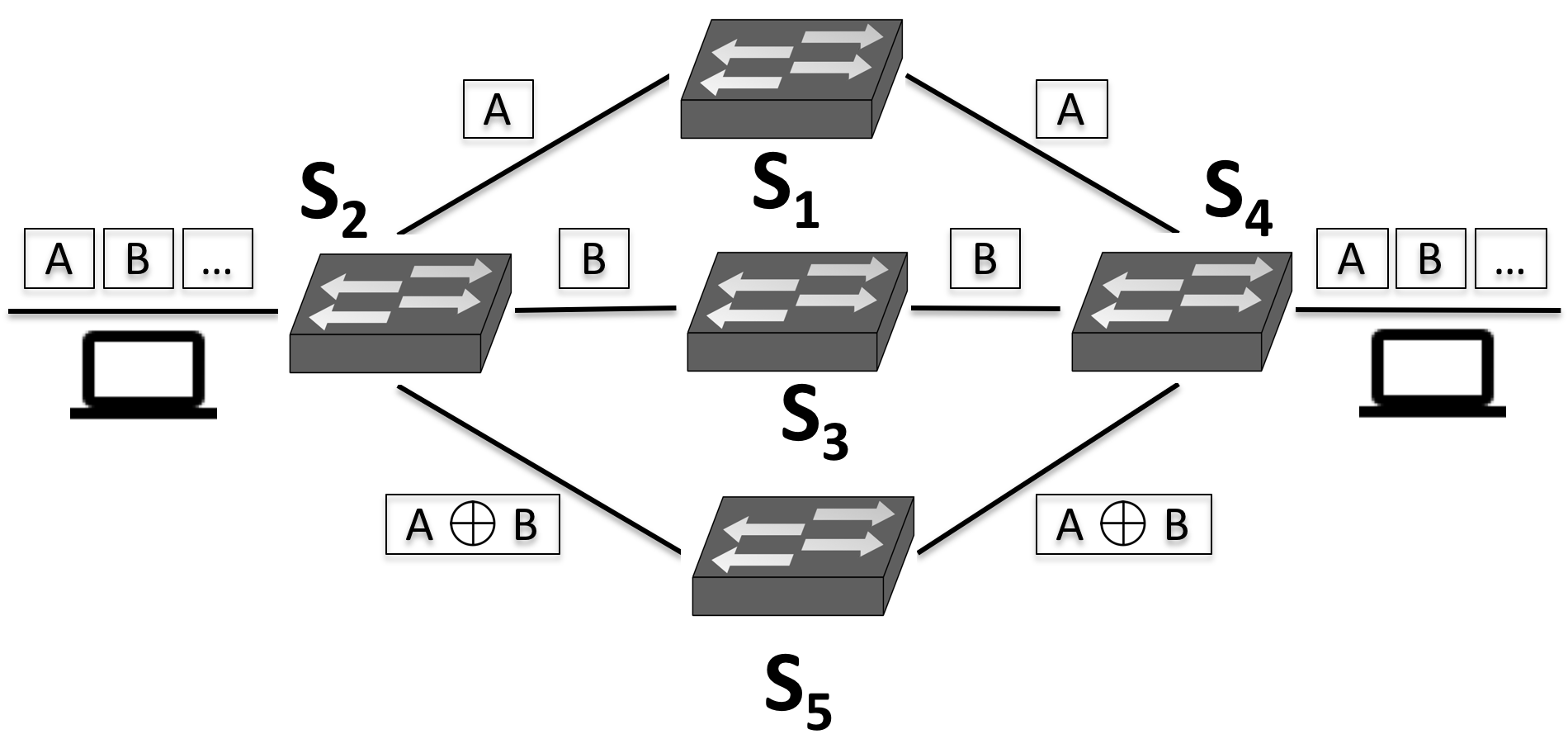}}\hfil}
\caption{Linear Coding Functions: (a) For multicast rate improvements over a butterfly topology (b) For seamless failure resilience to failure of links in any one of three available paths}
\label{fig:coding_functions}
\end{figure*}

\subsection{Coding Functions}
The coding functions are a part of the coding application. Each coding function takes as input the source host and destination host(s) associated with the data stream. It accesses the topology information by using the controller's northbound API. Then, the coding function generates the configuration for the coding primitives described later in this section. Figure \ref{fig:coding_functions} shows two instances of coding functions. One instance is that of a diversity code that provides seamless resilience for a unicast stream over three paths. The other instance uses a linear code for enhancing receiver's data rate of a multicast stream.

\subsection{Coding Header}
Each packet that belongs to a coding function carries a coding header. The header contains various fields to coordinate the operations performed by the coding primitives across the network. The header has a field called \texttt{next\_primitive} which determines what happens to the packet when it arrives at a network device. It also has a field called \texttt{stream\_id} to identify packets belonging to different streams. Finally, it has the \texttt{batch\_number} which identifies the packets belonging to a given batch of packets within the stream.

Since P4 does not provide access to the contents of a packet beyond a few hundred bytes, in our prototype, we use a field in the coding header to carry the packet's payload. 

\subsection{Coding Primitives}
The coding primitives are implemented in the data-plane on the individual network devices using P4.  The primitives are implemented primarily in the ingress pipeline. However, some of the primitives use the egress pipeline for recirculating cloned packets for generating new packets that carry coded/decoded payload. For every primitive, we also collect some in-band telemetry to measure processing times for evaluation. 

Coding primitives use several common design patterns. Each primitive uses at least one table in the ingress pipeline. If the primitive uses packet cloning and recirculation, then it also uses a table in the egress pipeline. Furthermore, each primitive table has a common field called \texttt{stream\_id} as part its \texttt{key}. This field is used to specify the packets belonging to an specific application's data stream. These packets could originate at the host or could be the result of the output of another primitive.

Each table implements a decision tree. The levels of the tree are determined by values taken by the fields in the \texttt{key} of the table. The actions in the table form the leaves in the decision tree. These actions either perform the mathematical operation for coding/decoding packets or manipulate some global state that is held in registers.

Next, we describe the particulars of the individual primitives. As illustrated in the Figure \ref{fig:primitives}, we developed five primitives to implement linear inter-session NC as follows:

\begin{itemize}
\item \textbf{Splitting:} primitive splits a given packet stream arriving from a single interface into individual batches of packets. It uses global state in registers on a per stream basis and stores the packets in the appropriate registers so that the coding/decoding primitives can use them.

\item \textbf{Coding:} primitive \textit{generates} new packets whose payload are obtained by coding over the previously stored payloads. It accomplishes that by using the cloning and recirculation features to create a loop to generate packets whose payload is then populated to be the coded packets. In order implement linear codes, the coding primitive only performs addition, subtraction and XOR operations provided by P4.

\item \textbf{Forwarding:} primitive performs unicast or multicast forwarding of a packet. The multicast forwarding action makes use of cloning to generate copies of packets.

\item \textbf{Gathering:} primitive collects a batch of incoming packets from multiple interfaces and puts them into the registers corresponding to their \texttt{stream\_id}. It also relies on global state in registers to keep track of the packets it has received on a per-stream basis.

\item \textbf{Decoding:} primitive takes the gathered packets, decodes them and forwards the payload packets to the host. In order to generate decoded payload, this operation may also require \textit{generating} new packets. This primitive also uses cloning and recirculation to generate new packets that are then populated with decoded payload.
\end{itemize}

Clearly, the coding and decoding primitives require buffering of the packets. In our current prototype, we use the registers to maintain a fixed size ring buffer containing the packet payloads and use the \texttt{batch\_number} to locate them appropriately. The packets belonging to same stream of packets are correlated by using the \texttt{stream\_id}. In our design, we do not perform any flow control on the switches.
\section{Evaluation}
\label{sec:evaluation}

We implemented a prototype of the library of primitives and functions that use them. Our prototype is available in the public domain \cite{open_source_aquaflow}. We evaluated our approach using \texttt{mininet} \cite{lantz2010network} and a software switch \cite{bmv2} as P4 target. The end-host were emulated using python scripts that used \texttt{scapy} \cite{scapy} to construct and parse custom coding headers. The emulations were performed on a machine that was running Ubuntu 16.04 LTS. The machine had eight processor cores clocked at 2.7 GHz and 16 GB of RAM. We performed two types of evaluation that are described below.

\begin{figure}
\centering
\includegraphics[width=3.5in]{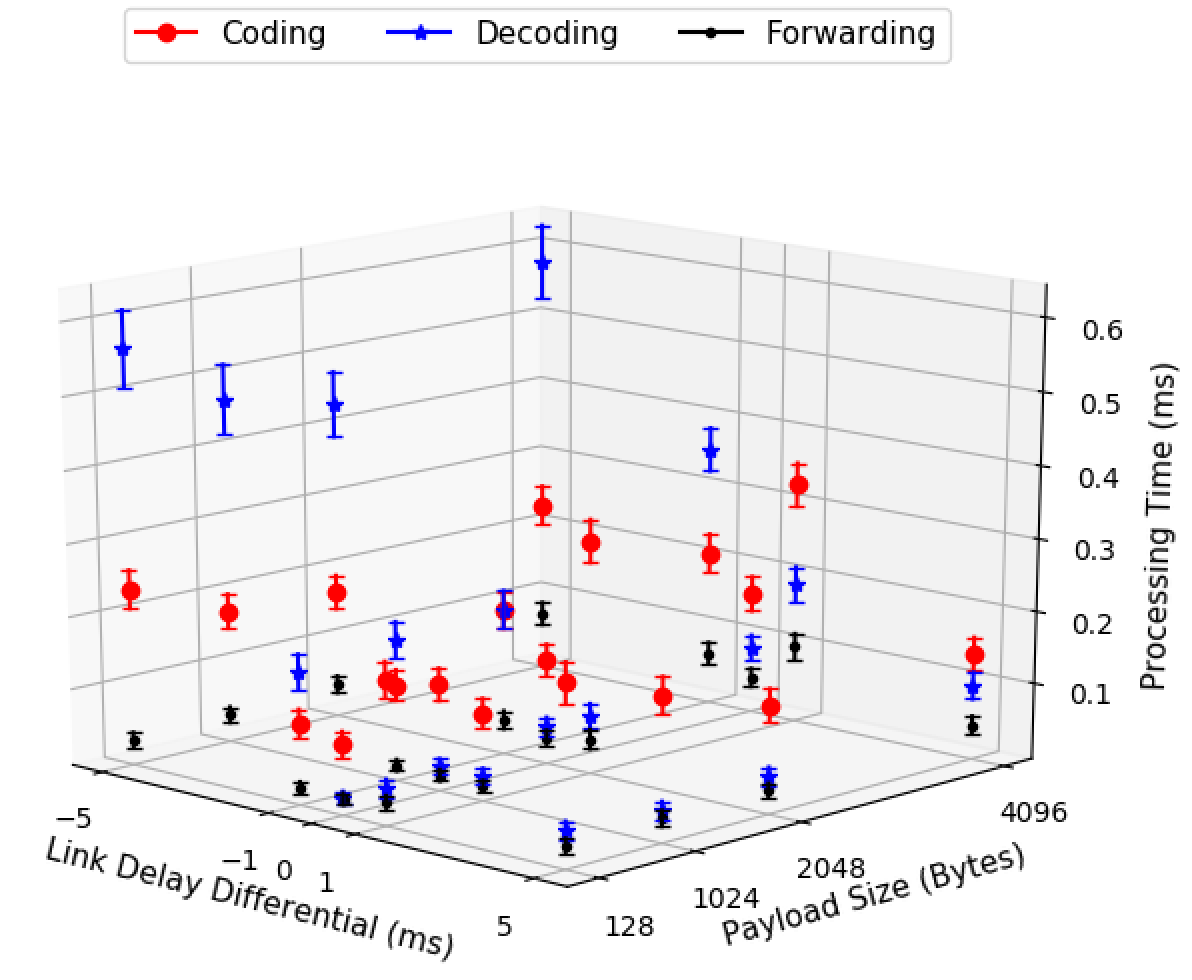}
\caption{Processing time per packet for coding and decoding using the diversity code}
\label{fig:diversity_experiment}
\end{figure} 

\subsection{Multicast Rate Gains of Coding}
We performed an experiment to measure the multicast rate gains that are obtained when using a simple linear code to perform multicast over the classic butterfly network shown in Figure \ref{fig:coding_functions}(a). The host on the left side wants to multicast a data stream to the two hosts on the right. Suppose the bandwidth of the links between the switches is $k$ bps. Theoretically, coding should allow a multicast rate of $k$ bps for flows $S_1 \rightarrow S_2$ and $S_1 \rightarrow S_3$ simultaneously, whereas any scheme that uses packet forwarding would not be able to accomplish this rate because only a single packet can be forwarded along the link $S_4 - S_5$.

\begin{figure}
\centering
\includegraphics[width=3.5in]{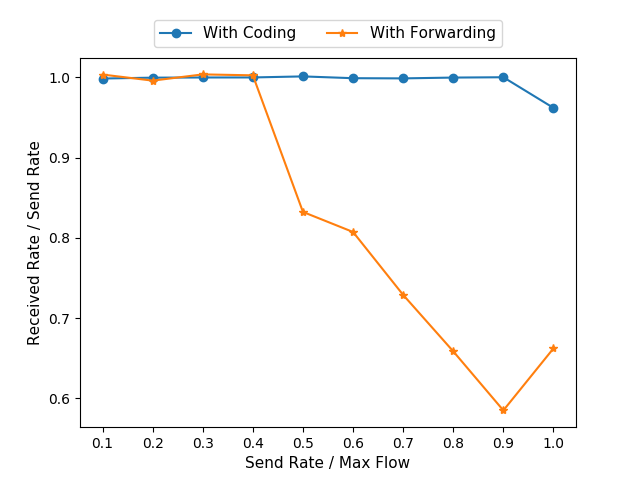}
\caption{Comparison between received rate achieved using coding vs forwarding in the butterfly topology}
\label{fig:butterfly_experiment}
\end{figure}

We sent a thousand packets at various data send rates with an exponentially distributed inter-packet time to match the rate. Each packet contained 4096 bytes of payload. The $k$ was set to $0.01$ Mbps. We measured the simultaneous received data rate on each receiver by using the packet payload sizes and their timestamps in the generated PCAP files to obtain the range of time for which the transmission was received at each receiver,

Figure \ref{fig:butterfly_experiment} plots two ratios. The x-axis is the ratio of send-rate over the max-flow between the source and destination host, whereas the y-axis is the ratio of observed received rate at one of the receiving hosts (without loss of generality and empirically identical) over the send rate. We observed that the received rate drops for forwarding as the send rate is about 40\% of the max-flow, whereas it does not drop for coding until 80\% of the max-flow. At 80\%, we see a drop for coding as well, likely because the processing delays associated with coding at nodes $S_4$, $S_6$, $S_7$.

\subsection{Microbenchmarks}
We performed an experiment to measure the processing time associated with coding and decoding packets in a P4-enabled switch. We used in-band telemetry to measure the processing time at each switch in the path of the packet. The processing time is measured as difference between the time-stamp associated with packet arriving at the ingress pipeline and packet being queued for egress.

We used the framework described above to implement diversity coding over multiple alternative paths as shown in Figure \ref{fig:coding_functions}(b). Each link was set to have a delay of 5 ms. The delay for link between $S_1$-$S_4$ was varied to create a delay differential for packets arriving at $S_5$ for decoding. Furthermore, we measured the impact of payload size on the processing time as well. We sent a thousand packets for a each point in the plot shown in Figure \ref{fig:diversity_experiment}. The packets were sent as fast as possible (i.e. there was no sleep between any two packets).

We observed a negligible effect of increasing the payload sizes of packets for all operations. Specifically, we observed that the processing time for the set of switches that forward the packets (i.e. $S_2$, $S_3$, $S_4$) increases only slightly even when payload sizes are quadrupled. In the worst case, we observe that processing time at the coding node (i.e. $S_1$) can be up to four times the processing time for only forwarding the packets. Similarly, in the worst-case, decoding node (i.e. $S_5$) can take up to six times to process a packet than a forwarding node. Some of this variation and extra processing time is due to the cloning and recirculation operation that coding and decoding primitives use. 

Finally, we observe that for a lower link delay differential, the decoding time is higher than coding time and has a high standard deviation and vice versa. This is because different set of table actions are in effect in those two cases. For a lower link delay differential, the XOR packet arrives at the decoder first and necessitates the use of arithmetic decoding, whereas when the differential is higher, the decoding is essentially reduced to forwarding the uncoded packets.

\section{Conclusion and Future Work}
\label{sec:conclusion_future_work}
We proposed an architecture that realizes linear NC in programmable networks by decomposing the linear NC functions into the atomic coding primitives. In our future work, we want to verify whether the gains associated with NC are also observed in hardware devices. Furthermore, we want to study the resilience and throughput gains for more complex coding functions and in larger topologies. Finally, we also want to study how NC can be used to control the trade-offs between network resources, the level of resilience and end-to-end delay.

\bibliographystyle{IEEEtran}
\bibliography{bibliography} 

\end{document}